# The Air–Water Interface of Condensed Water Microdroplets does not Produce $H_2O_2$

Nayara H. Musskopf[&], Adair Gallo Jr.[&], Peng Zhang[&], Jeferson Petry, Himanshu Mishra[*]

Interfacial Lab (iLab), King Abdullah University of Science and Technology (KAUST), Biological and Environmental Science and Engineering (BESE) Division, Water Desalination and Reuse Center (WDRC), Thuwal, 23955-6900, Saudi Arabia.

[&]Equal author contribution

[*]Correspondence: Himanshu.Mishra@kaust.edu.sa

**Abstract**

Recent reports on the production of hydrogen peroxide ($H_2O_2$) on the surface of condensed water microdroplets without the addition of catalysts or additives have sparked significant interest. The underlying mechanism is speculated to be ultrahigh electric fields at the air-water interface; smaller droplets present higher interfacial area and produce higher (detectable) $H_2O_2$ yields. Herein, we present an alternative explanation for these experimental observations. We compare $H_2O_2$ production in water microdroplets condensed from vapor produced via (i) heating water to 50–70 °C and (ii) ultrasonic humidification (as exploited in the original report). Water microdroplets condensed after heating do not show any enhancement in the $H_2O_2$ level in comparison to the bulk water, regardless of droplet size or the substrate wettability. In contrast, those condensed after ultrasonic humidification produce significantly higher $H_2O_2$ quantities. We conclude that the ultrasonication of water contributes to the $H_2O_2$ production, not droplet interfacial effects.

Recent studies reporting on the spontaneous production of ≤ 115 μM $H_2O_2$ in condensed water microdroplets of diameter ≤ 10 μm on common substrates have sparked considerable interest[1]. Elucidation of the breakage of water O–H covalent bonds under normal temperature and pressure (NTP, 293 K and 1 atm) without the use of a catalyst, energy (electrical or mechanical), or other chemicals, poses a challenge to our current understanding of water. These findings could advance our understanding of environmental processes, such as the oxidation of S(IV) species by $H_2O_2$ in the presence/absence of $Fe^{2+}$ or $Cu^{2+}$ ions, leading to the acidification of clouds, dew, fog, etc.[2-4] Greener approaches for $H_2O_2$ synthesis[5] and the rational development of hydrogen peroxide-based automated room disinfection in hospitals[6], water treatment[7], oral care[8], etc., are other exciting avenues. The mechanism speculated for $H_2O_2$ production in water microdroplets is based on the presence an ultrahigh electric field at the air-water interface (~$10^7$ V/cm), which drives the formation of ·OH radicals from $OH^-$ ions; these ·OH radicals combine to form $H_2O_2$[1, 9]. Currently, there is no theoretical explanation available for this intriguing phenomenon. We note that probing the air-water interface of sub-1nm dimensions is a daunting task and sometimes fraught with interpretational ambiguities and/or experimental artifacts[10-19].

Let us begin by briefly discussing the significance of the microscale of water droplets implicated in this phenomenon. Microscale enhances the air–water surface area; the higher the droplet surface area, the higher is the $H_2O_2$ production due to the putative surface electrical field[1]. For example, the surface area of a 1 ml water droplet at normal temperature and pressure (NTP: 293 K and 1 atm) of diameter, $D$ = 1.24 cm, if transformed into microdroplets of $d$ = 5 μm diameter, increases by a factor of $D/d$ = 2480. To summarize, the claim essentially is that $H_2O_2$ is produced even on the surface of a pail of water, but it remains undetected owing to the low air-water surface area and large volume of water, yielding ultralow concentrations[1]. The maximum reported $H_2O_2$ concentration in the water microdroplets condensed from the vapor within the relative humidity range 40–70% and the substrate (silica) temperature 3.5 °C is ~115 µM, when the droplet mean size was ≤ 10 µm[1]. This concentration translates to ~1 $H_2O_2$ molecule per ~483,092 water molecules, which renders standard spectroscopic fingerprinting techniques ineffective. As the condensed droplets grow over time, their surface-to-volume ratios drop, plummeting $H_2O_2$ production and causing further reduction in the measurable $H_2O_2$ concentration[1]. Thus, it is crucial to utilize ultrasensitive detection methods to investigate the factors and mechanisms underlying this phenomenon. Herein, we investigate $H_2O_2$ in condensed water microdroplets via an ultrasensitive method and put forth an alternate explanation for $H_2O_2$ formation in condensed water microdroplets.

In order to utilize the most sensitive analytical method for quantifying trace amounts of $H_2O_2$ in water, we compared the efficacy of several known methods[3]. For instance, titration with 0.1 M potassium titanium oxalate (PTO, $K_2TiO(C_2O_4)_2$) is the most common method that was also exploited in the original report[1]. It entails the measurement of the absorbance at 400 nm of samples that also contain PTO (Figure S1A). However, this method can at best detect ~10 µM $H_2O_2$ (or ~0.34 ppm) (Figure 1A). We also tested the efficacy of titration with terephthalic acid that yields 2-hydroxyterephthalic acid (HTA) on reaction with ·OH radicals furnished by $H_2O_2$[20]. However, this method also fails to resolve dissolved $H_2O_2$ concentrations below 5 µM (or ~ 0.17 ppm) (Figures 1B and S1B). Either method would, for instance, prove to be inadequate in providing high-resolution insight into the time-dependent decline in the $H_2O_2$ concentration in condensing drops as they get larger and more dilute. In response, we utilized the Hydrogen Peroxide Assay Kit (HPAK, ab138886, Abcam PLC, Cambridge, UK). This fluorometric kit contains (i) a proprietary peroxide indicator that upon oxidation with $H_2O_2$ produces fluorescence in the near-infrared region (674 nm wavelength) and (ii) the horseradish peroxidase (HRP) enzyme that catalyzes the oxidation reaction. The HRP's catalytic activity enhances the fluorescence signal significantly, facilitating a linear detection range for $H_2O_2$ from ~250 nM (or 8.5 ppb) to 10 μM (0.3 ppm) and rendering HPAK one of the most sensitive quantification methods for $H_2O_2$.[3, 21] Figure 1A-C compares the calibration plots for $H_2O_2$ titrations with PTO, terephthalic acid, and HPAK methods. With an approximately 40-times lower limit of detection than the PTO method, the HPAK assay would be able to pinpoint $H_2O_2$ concentrations inside condensing water droplets as their volumes increase with time (despite the dilution).

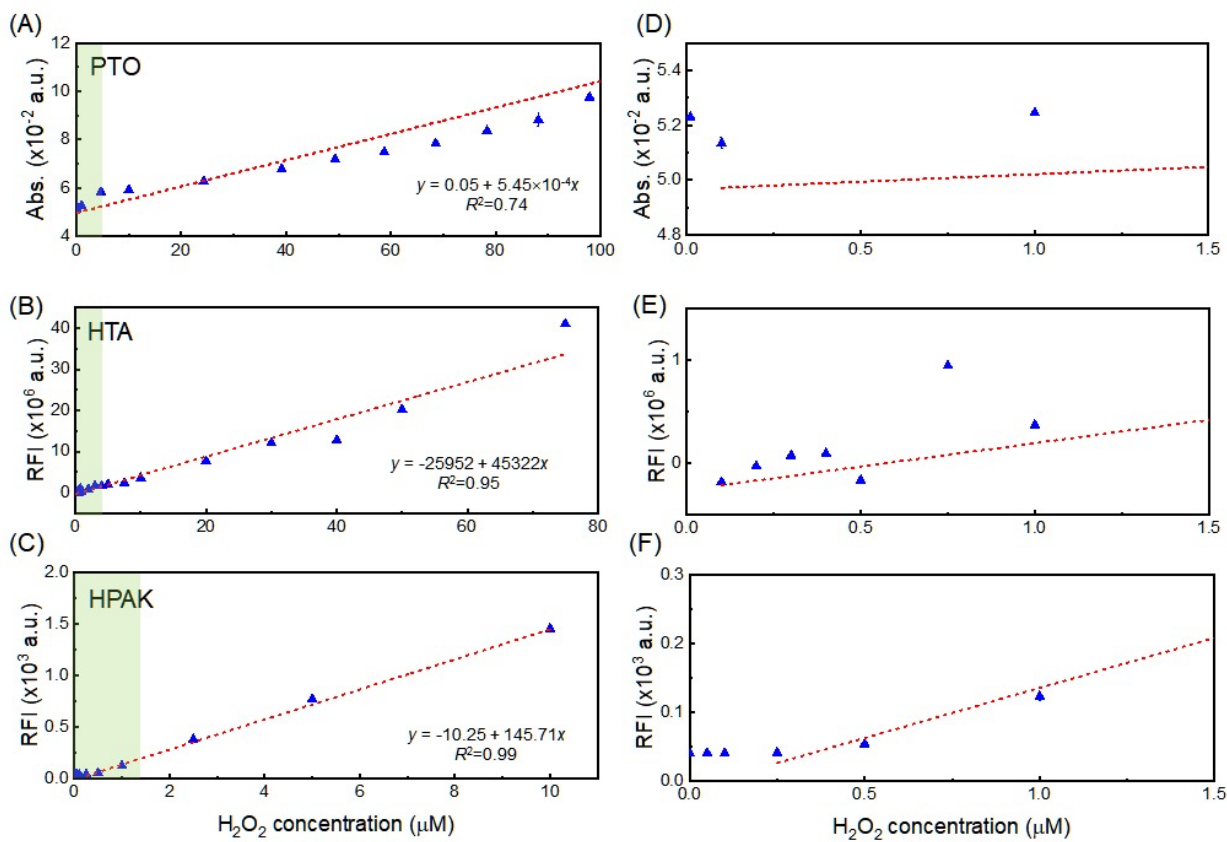

**Figure 1.** Comparison of representative calibration curves for $H_2O_2$ quantification assays. (A) the potassium titanium oxalate (PTO) assay, (B) the 2-hydroxyterephthalic acid (HTA) assay, and (C)

the hydrogen peroxide assay kit (HPAK). The HPAK can detect $H_2O_2$ concentrations down to ~250 nM range while the PTO and the HTA assays fail to resolve at 10 μM and 5 μM, respectively. In the 0–1 μM range, PTO (D) and HTA (E) assays cannot afford linearity in quantification, whereas HPAK (F) affords it reliably for ≥ 250 nM $H_2O_2$ concentrations.

Equipped with the HPAK assay, we investigated $H_2O_2$ production in condensed water microdroplets as a function of different: (i) methods for producing water vapor, (ii) substrates for condensation, and (iii) droplet size. Experiments were conducted inside a clean glovebox equipped with $N_2$ flow to control the relative humidity of air. Water vapor was produced inside the glove box via: (i) a commercial ultrasonic humidifier (15 W output power) equipped with a megahertz range piezoelectric transducer or (ii) a hot plate to heat water in the 50–70 °C range (Materials and Methods). N.B. an ultrasonic humidifier is a common household appliance; it produces tiny droplets at the air-water interface that rapidly evaporate to increase the relative humidity; a similar equipment was exploited in the original report[1]. In both cases, the relative humidity was adjusted to 92–96% at ~21–23 °C by a moisture controller by flowing $N_2$ gas into the glovebox whenever the humidity was above the setpoint. Single crystal $SiO_2$/Si wafers with the following surface treatment (wettability) served as the substrates for the condensation of water vapor: (i) oxygen plasma-treatment (hydrophilic) (ii) silanization with perfluorodecyltrichlorosilane (FDTS) using a molecular vapor deposition technique[22] (hydrophobic; details in the Materials and Methods section and wetting characterization in SI Section S1). Substrates were cooled below the dew point by placing them on a water-ice bag; the surface temperatures were measured using an infrared probe (Methods) and found to be in the range 3–4 °C. A significant distance (~40 cm) between the location of the substrates inside the glove box and the vapor source ensured that any airborne droplets/clusters would evaporate first to form vapor and then condense, as observed by the droplets' gradual increase in size (Figures 2 and S2). Note that hydrophobic FDTS-coated $SiO_2$/Si wafers (hereafter referred to as FDTS-coated silica) were utilized as the substrate for condensation in most of this work because they facilitated easier removal of the condensed water for further analysis than did hydrophilic plasma-treated silica.

With the water vapor supplied from the ultrasonic humidifier, size distributions (mean±standard deviation) of the condensed microdroplets on FDTS-coated silica shifted their peak positions from $10.3 \pm 3.2$ μm at 10 s to $103 \pm 98$ μm (bimodal distribution) at 20 min (Figures 2-3 and details in Materials and Methods). Similarly, with the vapor supplied from the heated water (60 °C), size distributions shifted their peak positions from $8.4 \pm 2.7$ μm at 10 s to $144 \pm 134$ μm (bimodal distribution) at 40 min (Figures S2-S3).

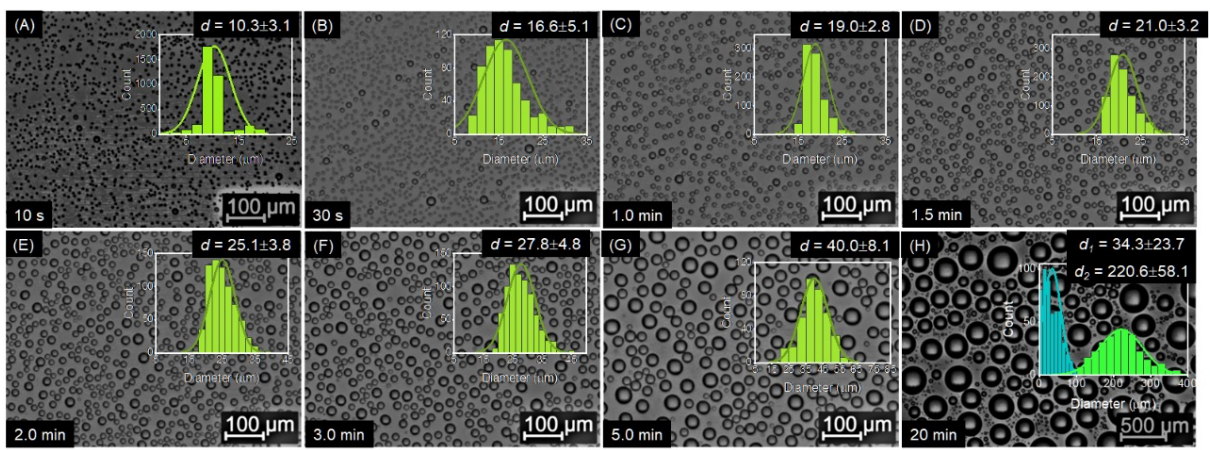

**Figure 2.** Representative time-dependent size distributions of condensed water droplets on FDTS-coated silica surfaces maintained at 3–4 °C and 92–96% RH realized via an ultrasonic humidifier: (A) 10 s, (B) 30 s, (C) 1 min, (D) 1.5 min, (E) 2.0 min, (F) 3.0 min, (G) 5 min, and (F) 40 min (bimodal distribution). Note: scale bars in μm.

We were also able to estimate the condensation rates via image analysis. To do this, we assumed the condensed microdroplets to be truncated spheres with the contact angles at the solid-liquid-vapor interface equal to the measured apparent contact angles on FDTS-coated silica, $\theta_r \approx 105°$ (SI Section S1). This is a reasonable assumption because the capillary length of water, defined as the length scale at which capillary forces dominate over inertia, at NTP is $\lambda_c = 2.7$ mm, which is much larger than the size of the condensed water microdroplets. N.B. $\lambda_c$ is given by the formula, $\lambda_c = \sqrt{\frac{\gamma_{LV}}{\rho g}}$, where, $\gamma_{LV}$ is the surface tension of water, $\rho$ is the density, and $g$ is the acceleration due to gravity[23]. With this approach, we estimated the condensation rates to be 0.31 µL/s and 0.23 µL/s for the ultrasonic humidifier and the heating plate (at 60 °C), respectively (Figures 3 and S3). Out of curiosity, we also quantified the condensation behavior of water from the ambient laboratory air at RH 59% and 21.4 °C and found it to be 0.35 µL/s (Figures S4-S5).

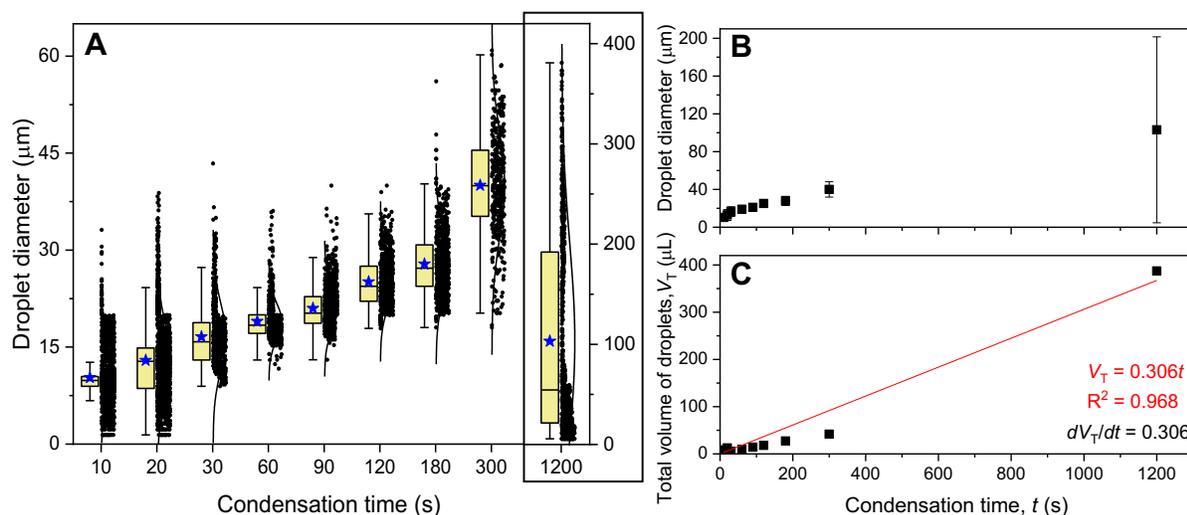

**Figure 3.** Representative image analysis (with ImageJ[24]) of condensed water droplets on FDTS-coated silica surfaces at 3–4 °C and 92–96% RH realized via an ultrasonic humidifier. (**A**) Condensed droplets' diameter distribution as a function of time (Note: blue stars correspond to the mean value; the x-axis scale is not linear and the right-hand-side y-axis is droplet diameter (µm) for $t$ = 1200 s only). (**B**) Mean droplet diameters as a function of time; error bars signify one standard deviation. (**C**) Estimated cumulative volume of the condensate over time; estimated average condensation rate: 0.31 µL/s. (Notice the bimodal droplet distribution at $t$ = 1200 s; this happens because larger droplets merge and numerous ≤ 10 µm droplets emerge in the spaces between them).

Next, $H_2O_2$ concentrations were quantified in the condensates formed on FDTS-coated silica and plasma-treated silica with the vapor formed via heating water or ultrasonic humidification (Figure 3). In a typical experiment, we stopped water condensation when the condensate was enough for us to collect ~400–600 µL, which took ~40 min for heating and ~20 min for the ultrasonic humidifier. See Figures 3A and S3A to notice the size distribution of droplets at those instances. Subsequently, the substrates were tilted to pour condensed water (drops or films) into clean glassware for $H_2O_2$ quantification via HPAK. When water was heated to 60 °C to produce the vapor, we found no significant differences (with a *p*-value > 0.01) in the $H_2O_2$ concentrations of the condensed water and the bulk water, i.e., both were below our detection limit (0.25 µM) (Figure 4 – group **c**, and Figure S8). Also, when the vapor was formed via heating water, substrates' hydrophilicity/hydrophobicity had no effect on the $H_2O_2$ concentration. These trends were consistent at 50 °C and 70 °C as well (Figure S6).

In contrast, when ultrasonic humidification was exploited to produce the vapor, there was a significant enhancement in the $H_2O_2$ concentration (~1 µM) in the condensed water (Figure 4); the condensates on the plasma-treated surfaces had a slightly higher $H_2O_2$ concentration than those on the FDTS-coated surfaces (groups **a** and **b** in Fig. 4).

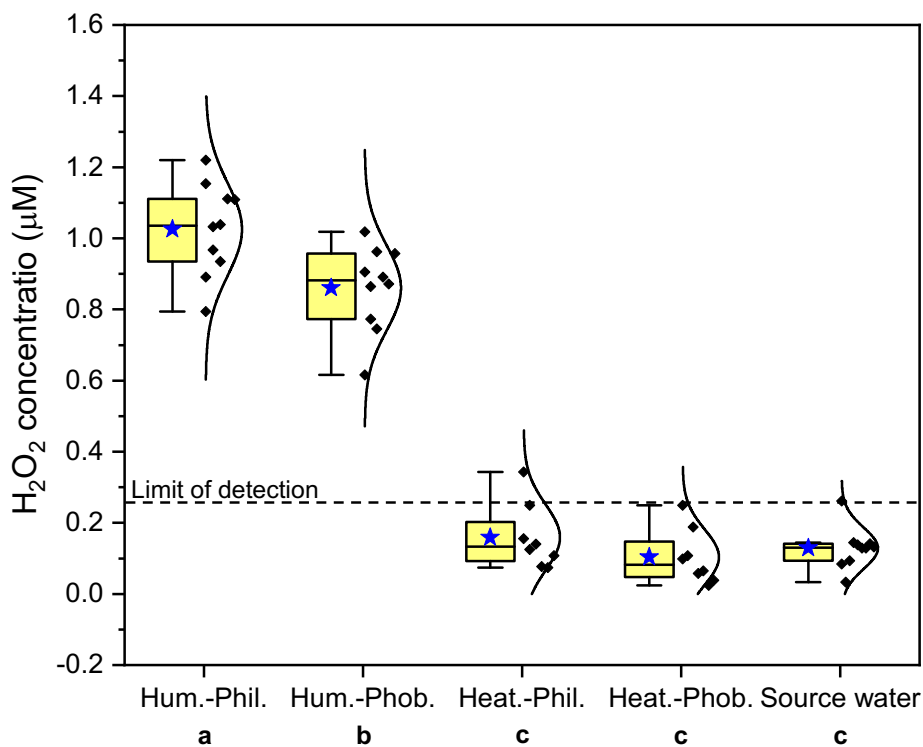

**Figure 4.** Comparison of $H_2O_2$ concentrations in the water microdroplets accumulated on hydrophobic (Phob.) and hydrophilic (Phil.) substrates after the condensation of vapor produced via (i) heating water in the temperature range of 50–70 °C (Heat.), and (ii) ultrasonic humidifier (Hum.). There were no statistically significant differences in the $H_2O_2$ concentration of the bulk water and the water droplets collected after the condensation of vapor generated by heating water at 60 °C. (Note: experimental results for heating at 50 °C and 70 °C (Fig. S6) yield results similar to the 60 °C case presented here). The chemical make-up of the substrates had no significant effects on the $H_2O_2$ concentration. In contrast, the $H_2O_2$ concentration in the water condensed from the ultrasonic humidifier was ~1 µM. In this case, the chemical make-up of the substrates also had a significant effect on the $H_2O_2$ concentration. Bold letters (**a**, **b** and **c**) under the labels refer to the statistically different groups analyzed with one-way ANOVA and Tukey's test for comparison of the means ($p < 0.01$); blue star indicates the mean value. (Note: the detection limit of the HPAK assay in our experiments was 0.25 µM).

We wondered if the absence of $H_2O_2$ in the condensates derived from the heating experiments and the ~1 µM $H_2O_2$ concentration measured in the condensates derived from the ultrasonic humidifier were so low due to the fact that:

(i) we collected microdroplets largely with diameters ≥ 10 µm, which caused a dramatic dilution of the $H_2O_2$ concentration; or

(ii)    $H_2O_2$ is produced exclusively when ultrasonic humidification is applied.

To pinpoint the correct answer, first, we tried to measure $H_2O_2$ concentration in water microdroplets of diameters strictly ≤ 10 μm. Under our experimental conditions, condensed water microdroplets' with diameters ≤10 μm appeared during the first 10 s for the ultrasonic humidifier (Figures 2-3) and during the first 20 s for the heating plate at 60 °C (Figures S2-S3) and the ambient moisture condensation (Figures S4-S5). Very quickly, we realized that due to their significantly smaller size than the capillary length of water, pinning forces were so high compared to inertia that it was not possible to detach them via tilting (90° or 180°) from the FDTS-coated silica substrates[25]. Microdroplets also evaporated rapidly when exposed to a lower relative humidity environment, which prevented us from using ultracentrifugation for sliding and collecting them. Furthermore, the volume of the droplets after only 10–20 s of condensation was so low that HPAK could not be utilized because the minimum sample volume, considering evaporative losses during handling, is 400–600 μL. Therefore, we resorted to commercial peroxide test strips with a detection limit of 29.4 μM for aqueous $H_2O_2$; similar strips were also utilized in the original report[1] (details in Materials and Methods). However, when peroxide test strips when swiped over water microdroplets of diameters ≤ 10 μm condensed from vapor generated by the ultrasonic humidifier, heating (60 °C), and ambient air, they did not undergo any change in their color. Note: as a check, we tested the test strips with standard $H_2O_2$ solutions and confirmed that they reliably detect $H_2O_2$ at concentrations ≥30 μM.

We wondered if this outcome was due to the fact that the lifetime of the microdroplets was too short to produce detectable $H_2O_2$ concentration. After the first few seconds (7-10 s) of condensation from the three sources listed above, we cut off the vapor from condensing further by placing a lid on top. This simple method allowed us to maintain the desirable size distribution for a longer time (2–5 minutes). Using peroxide test strips again, we found no evidence for aqueous $H_2O_2$ regardless of the waiting time.

Next, we studied the effects of the output power of ultrasonic humidifiers on the $H_2O_2$ concentration in condensed water microdroplets. We compared the performance of our 15 W ultrasonic humidifier with that of another commercial 20 W device (details in Materials and Methods). We broadened the investigation by probing $H_2O_2$ concentrations not only in the condensed microdroplets, but also in the mist formed at the humidifier outlet as well as the water reservoirs inside the humidifiers. Since the volume of the mist and the water reservoirs were large, we could track them from inception via HPAK, while we had to wait for ~20 min to collect adequate condensate (also explained above). Experimental results revealed that for either device, the $H_2O_2$ concentration was below detection limit before it was turned on; after mist formation started, the $H_2O_2$ concentration in the mist increased with time and reached ~2 μM and ~3 μM for the 15W and 20W devices, respectively (Figure 5). $H_2O_2$ concentrations in the reservoir also followed the same trend.

We also tested the effects of a 500 W microtip ultrasonication device (20,000 Hz) on $H_2O_2$ concentration in bulk water (Methods). Analysis with HPAK revealed that while the bulk water had an initial $H_2O_2$ concentration below the detection limit, it increased linearly with time to 3 μM in 3 hours (Figure S7).

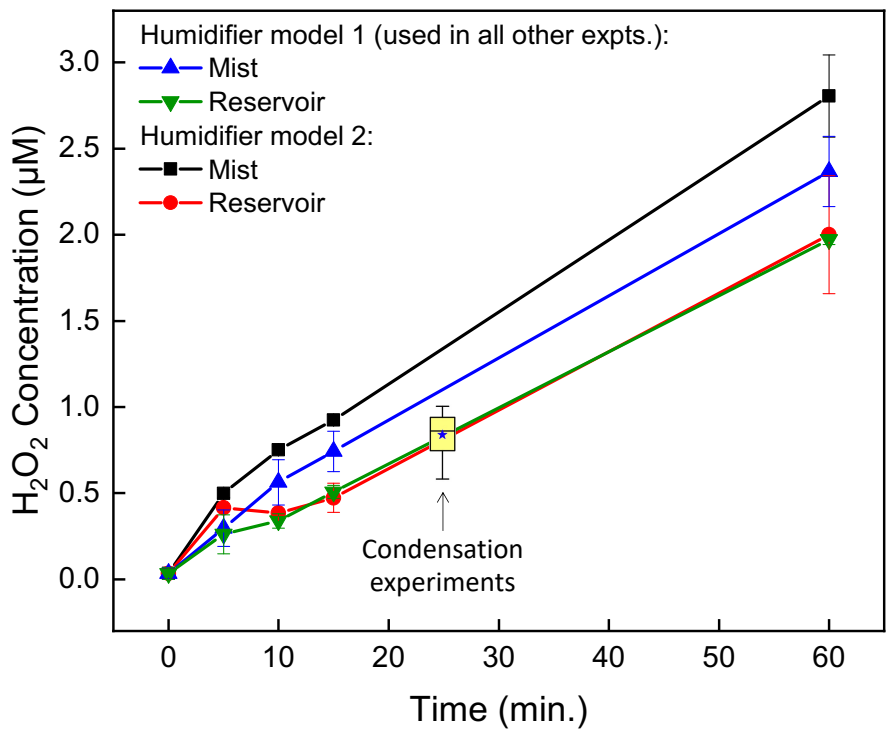

**Figure 5.** Ultrasonic humidifier mist and $H_2O_2$ formation. $H_2O_2$ formation over time in the mist and in the water reservoir (in contact with the piezo element) in ultrasonic humidifier models 1 and 2 with output powers of 15 W and 20 W, respectively. The yellow box represents the $H_2O_2$ concentrations in water microdroplets accumulated on FDTS-coated silica (from Figure 4). Note: Humidifier model 1 was used in most of the condensation experiments reported in this work and Humidifier model 2 was used to compare the effect of output power on $H_2O_2$ production.

As illustrated in Figure 6, our experiments with heating, ultrasonic humidifiers, and ambient air condensation demonstrate that detectable $H_2O_2$ is produced in condensed water microdroplets exclusively when ultrasonic humidifiers are exploited. In SI Section S2, we explain why condensed microdroplets with ≤10μm diameter produced via heating/ultrasonics/ambient air could not have had $H_2O_2$ concentrations ≥60 μM in our experiments, or else HPAK would have detected it. Furthermore, during ultrasonic humidification studies, we noticed that most of the generated mist fell back into the water reservoir, which caused the time-dependent rise in the $H_2O_2$ concentration in the mist. This is expected to increase the $H_2O_2$ concentration in the condensate for some time – a trend also observed in the original report[1]. For further insight into $H_2O_2$ production in ultrasonic humidifiers, it is crucial to investigate the effects of output power, operational frequency, structural design, and water level, etc., factors. How $H_2O_2$ partitions between the mist and the bulk water considering its ultrahigh solubility in water is also worth exploring. We continue to wonder about the mechanisms that led ≤115 µM level $H_2O_2$ concentrations in the latest report[1] that we simply cannot reproduce; we also wonder how the researchers overcame the solid–liquid adhesion to collect condensed droplets with diameters ≤10µm for the PTO analysis.

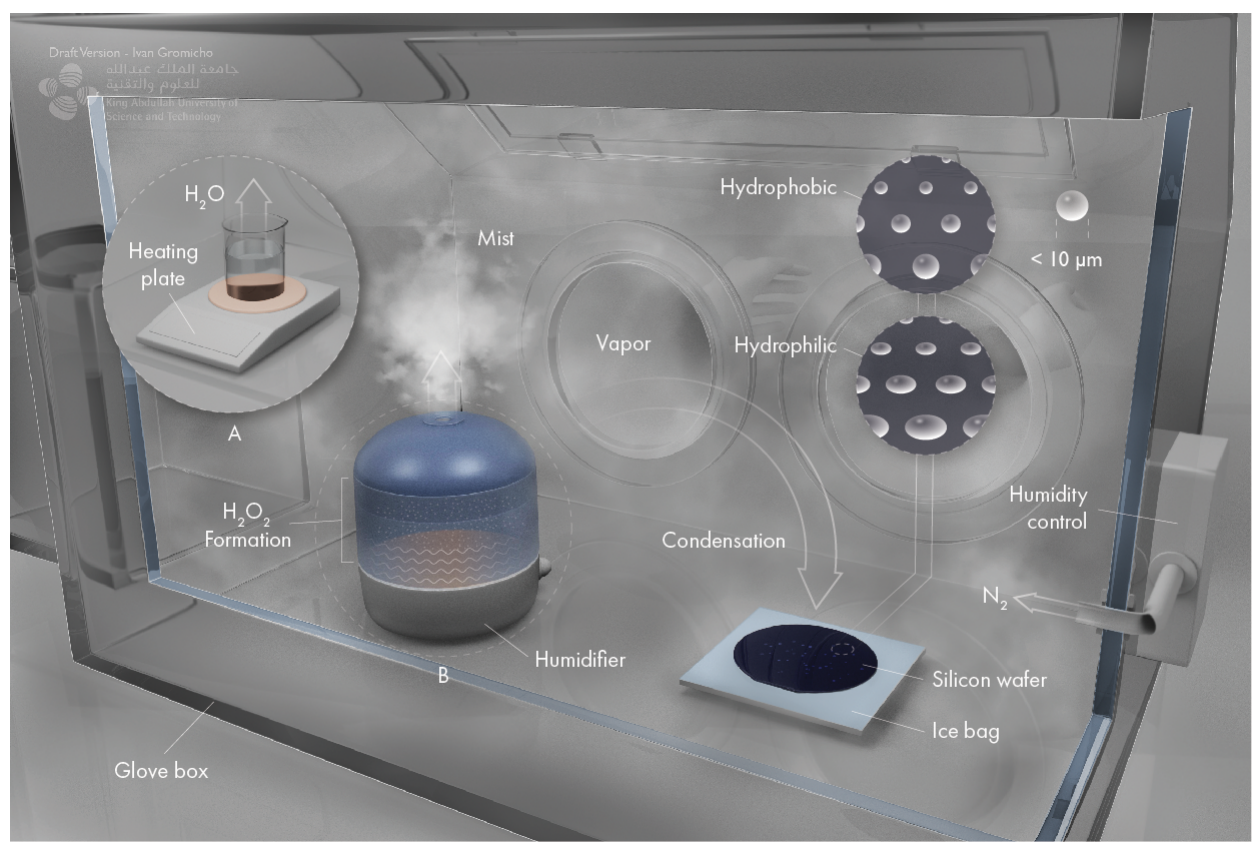

**Figure 6.** Scientific illustration of condensation of water vapor generated via (**A**) heating water on a hot plate and (**B**) an ultrasonic humidifier. Condensed water droplets on hydrophilic and hydrophobic substrates and then collected and analyzed for $H_2O_2$ concentration. Our experimental results reveal that the ultrasonic humidifiers can produce significant quantities of $H_2O_2$ in their mist (and also in the water reservoir), which then appears in the condensed water droplets. On the other hand, heating water to form vapor and condensing it does not produce $H_2O_2$ within our detection limits (≥0.25 µM). (Image credits: Ivan Gromicho, KAUST)

Significantly different outcomes manifest when a bulk liquid is exposed to ultrasonication or gently heated on a hot plate (below its boiling point and with a stirrer). The latter method distributes the energy uniformly within the bulk, while the former entails energy localization leading to hot-spots[26]. The ultrasonication of water causes the formation of numerous tiny pulsating bubbles – cavitation – that eventually implode to produce ultrahigh pressure (shock waves) and temperature and drive chemical reactions (including $H_2O_2$ formation)[27-29]. On the one hand, cavitation erosion of surfaces poses engineering challenge to fluid machinery[30, 31], on the other it has been utilized for wastewater treatment[32, 7, 33, 34], surface cleaning[35], and materials synthesis[36]. Similarly, ultrasonic humidifiers exploit MHz-frequency piezoelectric transducers that convert electrical energy into vibrational energy to produce the mist. Our experimental results demonstrate that this process produces $H_2O_2$ in the mist and, consequently, in the condensed droplets (Figure 5). In contrast, mild heating does not lead to $H_2O_2$ formation (Figure 4). Therefore, we submit that the recently reported $H_2O_2$ production in condensed water droplets could be in part due to experimental artifacts arising from ultrasonication[1]. By deduction, our findings also challenge the speculated mechanism for the $H_2O_2$ formation in (electrically neutral) water microdroplets due to the presence of ultrahigh electric fields (~$10^7$ V/m) at the air-water interface. Finally, these findings also establish that the air–water interface of condensed water microdroplets in environmental sources such as clouds, fog, dew, mist, and rain, etc., do not contribute to $H_2O_2$ formation due to interfacial effects.

## Materials and Methods

### Chemicals
A MilliQ Advantage 10 set up was utilized for deionized water used in this study. The water purification unit consisted of a Q-Gard pretreatment pack, UV lamp, Quantum cartridge (activated carbon and ion exchange resins) and a Q-Pod dispenser for final polishing[37]. The electrical resistivity of the water was 18.2 MΩ-cm. Standard hydrogen peroxide ($H_2O_2$) 30% solutions were purchased from VWR Chemicals (catalogue 23622.298) as used as-is.

### Substrates for Condensation
Silicon wafers (p-doped, <100> orientation, 4" diameter, with 500 μm thickness and with a thermally grown 2 μm-thick oxide layer) were purchased from Silicon Valley Microelectronics (Catalogue# SV010).

### Functionalization of $SiO_2$/Si wafers
**Oxygen plasma treatment for hydrophilic substrates:** Diener electronic machine (Atto model, 200 W) was supplied with ultrapure (99.999%) $O_2$ gas at the flow rate of 16.5 sccm for 10 min, to create oxygen plasma.

**Silanization with perfluorodecyltrichlorosilane (FDTS):** Treatment of silica surfaces with an oxygen plasma for 2 min removed organic contaminants and hydroxylated the surface. These surfaces were then grafted with FDTS using a molecular vapor deposition process (Applied Microstructures MVD100E) using protocols that we have reported previously[38].

### Characterizing Wetting of Substrates via Apparent Contact Angles
Apparent contact angles of water droplets were measured on the substrates using the Kruss Drop Shape Analyzer (DSA100E) and analyzed the data with the *Advance* software.

### Glove-box Experiments
Condensation experiments were performed inside a portable isolation glovebox (Cleatech, Catalogue#2200-2-B) as a controlled-environment chamber. It was equipped with a digital humidity control system (Cleatech, A21-HM-HDS) that purged nitrogen gas flow to dehumidify the air. $SiO_2$/Si wafers with the abovementioned surface treatments were used as substrates. The substrates were cooled down by placing them onto an ice–water bag, i.e., ice mixed with water such that the temperature of the bag is constant in every region. We allowed the surfaces to reach thermal equilibrium with the ice-water bath. The temperature of the surfaces was measured via a non-contact digital infrared thermometer (Lasergrip 774) during the experiment. The relative humidity inside the chamber was kept in the range 92–96% and the laboratory temperature was in the range 21–23 °C. To collect the samples after condensation, we poured the droplets into a clean glassware followed by transfer into a 15 mL centrifuge tube (VWR International).

**Water Vapor Generation via Ultrasonic Humidifier**

The following two ultrasonic humidifiers were used in this work: (i) Proton PHC 9UH (15W) and (ii) Beurer LB 44 (20W). Both of them contain a piezoelectric disk that vibrates creating ultrasonic waves leading to the formation of mist from bulk water. The former humidifier was used in most of our experiments. To prevent the direct deposition of the mist (before its evaporation) onto the substrates, we positioned them 40 cm apart. These humidifiers enabled us to pinpoint the effect of output power on $H_2O_2$ production in the mist, the water reservoirs, and the condensates.

**Water Vapor Generation via a Heating Plate**

Deionized water was heated in the 50–70 °C range using an IKA RCT heating plate (catalogue 3810000). The plate was located ~30 cm away from the substrates. To control the temperature, the coupled temperature sensor (PT 1000.60) was inserted inside the water.

**Quantification of $H_2O_2$ in water**

**Hydrogen Peroxide Assay Kit (HPAK) assay**

$H_2O_2$ concentration inside condensed water microdroplets was quantified using the Hydrogen Peroxide Assay Kit (Fluorometric-Near Infrared, Catalogue # ab138886). It contains its unique AbIR Peroxidase Indicator that produces fluorescence independent of the solution pH in the range 4–10. Its maximum excitation wavelength is at 647 nm and maximum emission at 674 nm. Horseradish peroxidase enzyme catalyzes the reaction between $H_2O_2$ and the indicator and enhances the fluorescence signal. This facilitates the linear range of detection from 250 nM to 10 µM. The calibration curve (Figure 1C) was realized by adding 50 µL of an $H_2O_2$ standard solution from a concentration of 50 nM to 10 µM into 50 µL of the $H_2O_2$ reaction mixture. The analysis was performed in a 96-well black/clear bottom microtiter-plate with the SpectraMax M3 microplate reader (Molecular Devices LLC). The analysis software used was SoftMax Pro 7. The water microdroplets were analyzed the same way by mixing 50 µL of each sample with the $H_2O_2$ reaction mixture, thus obtaining the respective concentration by the calibration curve.

**Peroxide test strips for semi-quantitative analysis:**

Peroxide test strips (Baker Test Strips, VWR International) with a detection limit of 1-100 mg/L were used to analyze $H_2O_2$ concentration when we could not collect adequate sample volumes (100 µL) for HPAK analysis. These strips contain a colorimetric reagent that turns blue when brought in contact with $H_2O_2$ in the specified concentration range.

**Microdroplets' size characterization:**

After the microdroplet condensation inside the chamber, the samples were quickly (< 10 s) moved to a Leica DVM6 optical microscope for imaging. Next, ImageJ software[24] was used to estimate the size distribution. For estimating the size distributions of microdroplets condensed from the ambient laboratory air, substrates were already positioned in the microscope.

**Probe sonication**

A 500 W ultrasonic processor (Sonics & Materials, Model VC 505) with a stepped microtip was used to ultrasonicate 30 ml water at 20 kHz at its 40% amplitude for different durations between 10–180 min. The beaker was kept in an ice bath to prevent sample evaporation.

# References

1. Lee, J. K.; Han, H. S.; Chaikasetsin, S.; Marron, D. P.; Waymouth, R. M.; Prinz, F. B.; Zare, R. N., Condensing water vapor to droplets generates hydrogen peroxide. *P Natl Acad Sci USA* **2020,** *117* (49), 30934-30941.
2. Seinfeld, J. H.; Pandis, S. N., *Atmospheric Chemistry and Physics: From Air Pollution to Climate Change*. Wiley-Interscience; 2 edition (August 11, 2006): 1998.
3. Gunz, D. W.; Hoffmann, M. R., Atmospheric chemistry of peroxides: a review. *Atmospheric Environment. Part A. General Topics* **1990,** *24* (7), 1601-1633.
4. He, S. Z.; Chen, Z. M.; Zhang, X.; Zhao, Y.; Huang, D. M.; Zhao, J. N.; Zhu, T.; Hu, M.; Zeng, L. M., Measurement of atmospheric hydrogen peroxide and organic peroxides in Beijing before and during the 2008 Olympic Games: Chemical and physical factors influencing their concentrations. *Journal of Geophysical Research: Atmospheres* **2010,** *115* (D17).
5. Zhu, C.; Francisco, J. S., Production of hydrogen peroxide enabled by microdroplets. *Proceedings of the National Academy of Sciences* **2019,** *116* (39), 19222.
6. Otter, J. A.; Yezli, S.; Barbut, F.; Perl, T. M., An overview of automated room disinfection systems: When to use them and how to choose them. *Decontamination in Hospitals and Healthcare* **2020**, 323-369.
7. Hoffmann, M.; Hua, I.; Hoechemer, R., Application of Ultrasonic Irradiation for the Degradation of Chemical Contaminants in Water. *Ultrasonics Sonochemistry* **1996,** *3*, 168-172.
8. Marshall, M.; Cancro, L.; Fischman, S., Hydrogen Peroxide: A Review of Its Use in Dentistry. *Journal of periodontology* **1995,** *66*, 786-96.
9. Dulay, M. T.; Huerta-Aguilar, C. A.; Chamberlayne, C. F.; Zare, R. N.; Davidse, A.; Vukovic, S., Effect of relative humidity on hydrogen peroxide production in water droplets. *QRB Discovery* **2021,** *2*, e8.
10. Mishra, H.; Enami, S.; Nielsen, R. J.; Stewart, L. A.; Hoffmann, M. R.; Goddard, W. A.; Colussi, A. J., Bronsted basicity of the air-water interface. *P Natl Acad Sci USA* **2012,** *109* (46), 18679-18683.
11. Saykally, R. J., Air/water interface: Two sides of the acid-base story. *Nature Chemistry* **2013,** *5* (2), 82-84.
12. Gallo, A.; Farinha, A. S. F.; Dinis, M.; Emwas, A.-H.; Santana, A.; Nielsen, R. J.; Goddard, W. A.; Mishra, H., The chemical reactions in electrosprays of water do not always correspond to those at the pristine air–water interface. *Chemical Science* **2019,** *10* (9), 2566-2577.
13. Colussi, A. J.; Enami, S., Comment on "The chemical reactions in electrosprays of water do not always correspond to those at the pristine air–water interface" by A. Gallo Jr, A. S. F. Farinha, M. Dinis, A.-H. Emwas, A. Santana, R. J. Nielsen, W. A. Goddard III and H. Mishra, Chem. Sci., 2019, 10, 2566. *Chemical Science* **2019**.
14. Gallo, A.; Farinha, A. S. F.; Emwas, A.-H.; Santana, A.; Nielsen, R. J.; Goddard, W. A.; Mishra, H., Reply to the 'Comment on "The chemical reactions in electrosprays of water do not always correspond to those at the pristine air–water interface"' by A. J. Colussi and S. Enami, Chem. Sci., 2019, 10, DOI: 10.1039/c9sc00991d. *Chemical Science* **2019**.
15. Nauruzbayeva, J.; Sun, Z.; Gallo, A.; Ibrahim, M.; Santamarina, J. C.; Mishra, H., Electrification at water–hydrophobe interfaces. *Nature Communications* **2020,** *11* (1), 5285.
16. Uematsu, Y.; Bonthuis, D. J.; Netz, R. R., Charged Surface-Active Impurities at Nanomolar Concentration Induce Jones-Ray Effect. *Journal of Physical Chemistry Letters* **2018,** *9* (1), 189-193.

17. Byrnes, S. J.; Geissler, P. L.; Shen, Y. R., Ambiguities in surface nonlinear spectroscopy calculations. *Chemical Physics Letters* **2011,** *516* (4-6), 115-124.
18. Agmon, N.; Bakker, H. J.; Campen, R. K.; Henchman, R. H.; Pohl, P.; Roke, S.; Thämer, M.; Hassanali, A., Protons and Hydroxide Ions in Aqueous Systems. *Chemical Reviews* **2016,** *116* (13), 7642-7672.
19. Ruiz-Lopez, M. F.; Francisco, J. S.; Martins-Costa, M. T. C.; Anglada, J. M., Molecular reactions at aqueous interfaces. *Nature Reviews Chemistry* **2020,** *4* (9), 459-475.
20. Barreto, J. C.; Smith, G. S.; Strobel, N. H.; McQuillin, P. A.; Miller, T. A., Terephthalic acid: a dosimeter for the detection of hydroxyl radicals in vitro. *Life sciences* **1994,** *56* (4), PL89-PL96.
21. Kim, G.; Lee, Y.-E. K.; Kopelman, R., Hydrogen Peroxide ($H_2O_2$) Detection with Nanoprobes for Biological Applications: A Mini-review. In *Oxidative Stress and Nanotechnology: Methods and Protocols*, Armstrong, D.; Bharali, D. J., Eds. Humana Press: Totowa, NJ, 2013; pp 101-114.
22. Pillai, S.; Santana, A.; Das, R.; Shrestha, B. R.; Manalastas, E.; Mishra, H., A molecular to macro level assessment of direct contact membrane distillation for separating organics from water. *Journal of Membrane Science* **2020,** *608*, 118140.
23. Adamson, A. W.; Gast, A. P., *Physical Chemistry of Surfaces*. Wiley-Interscience: 1997.
24. Schneider, C. A.; Rasband, W. S.; Eliceiri, K. W., NIH Image to ImageJ: 25 years of image analysis. *Nature Methods* **2012,** *9* (7), 671-675.
25. Gao, N.; Geyer, F.; Pilat, D.; Wooh, S.; doris, v.; Butt, H.-J.; Berger, R., How drops start sliding over solid surfaces. *Nature Physics* **2018,** *14*.
26. Suslick, K. S., Sonochemistry. *Science* **1990,** *247* (4949), 1439-45.
27. Riesz, P.; Berdahl, D.; Christman, C. L., Free radical generation by ultrasound in aqueous and nonaqueous solutions. *Environ Health Perspect* **1985,** *64*, 233-252.
28. Suslick, K. S., *Ultrasound: its chemical, physical, and biological effects*. VCH Publishers: 1988.
29. Fang, X.; Mark, G.; von Sonntag, C., OH radical formation by ultrasound in aqueous solutions Part I: the chemistry underlying the terephthalate dosimeter. *Ultrasonics Sonochemistry* **1996,** *3* (1), 57-63.
30. Gonzalez-Avila, S. R.; Nguyen, D. M.; Arunachalam, S.; Domingues, E. M.; Mishra, H.; Ohl, C.-D., Mitigating cavitation erosion using biomimetic gas-entrapping microtextured surfaces (GEMS). *Science Advances* **2020,** *6* (13), eaax6192.
31. Singh, R.; Tiwari, S.; Mishra, S., Cavitation Erosion in Hydraulic Turbine Components and Mitigation by Coatings: Current Status and Future Needs. *Journal of Materials Engineering and Performance* **2011,** *21*.
32. Gogate, P.; Tayal, R. K.; Pandit, A., Cavitation: A technology on the horizon. *Current Science* **2006,** *91*, 35-46.
33. Petrier, C.; Lamy, M.-F.; Francony, A.; Benahcene, A.; David, B.; Renaudin, V.; Gondrexon, N., Sonochemical Degradation of Phenol in Dilute Aqueous Solutions: Comparison of the Reaction Rates at 20 and 487 kHz. *The Journal of Physical Chemistry* **1994,** *98* (41), 10514-10520.
34. Hoffmann, M. R.; Hua, I.; Höchemer, R., Application of ultrasonic irradiation for the degradation of chemical contaminants in water. *Ultrasonics Sonochemistry* **1996,** *3* (3), S163-S172.

35. Ohl, C.-D.; Arora, M.; Dijkink, R.; Janve, V.; Lohse, D., Surface cleaning from laser-induced cavitation bubbles. *Applied Physics Letters* **2006,** *89* (7), 074102.
36. Bang, J. H.; Suslick, K. S., Applications of Ultrasound to the Synthesis of Nanostructured Materials. *Advanced Materials* **2010,** *22* (10), 1039-1059.
37. Millipore, M., Milli-Q® Advantage A10® Water Purification Systems. User-adapted ultrapure water. EMD Millipore Corporation, Billerica, MA, U.S.A.: 2013.
38. Shrestha, B. R.; Pillai, S.; Santana, A.; Donaldson Jr, S. H.; Pascal, T. A.; Mishra, H., Nuclear Quantum Effects in Hydrophobic Nanoconfinement. *The Journal of Physical Chemistry Letters* **2019,** *10* (18), 5530-5535.

**Acknowledgements**: The co-authors acknowledge research funding from King Abdullah University of Science and Technology under award number BAS/1/1070-01-01. Illustration for Figure 6 was created by Ivan Gromicho, Scientific Illustrator, Research Communication and Publication Services, Office of the Vice President for Research, King Abdullah University of Science and Technology.

# Supplementary Information

## The Air–Water Interface of Condensed Water Microdroplets does not Produce $H_2O_2$

Nayara H. Musskopf[&], Adair Gallo Jr.[&], Peng Zhang[&], Jeferson Petry, Himanshu Mishra[*]

Interfacial Lab (iLab), King Abdullah University of Science and Technology (KAUST), Biological and Environmental Science and Engineering (BESE) Division, Water Desalination and Reuse Center (WDRC), Thuwal, 23955-6900, Saudi Arabia.

[&]Equal author contribution

[*]Correspondence: Himanshu.Mishra@kaust.edu.sa

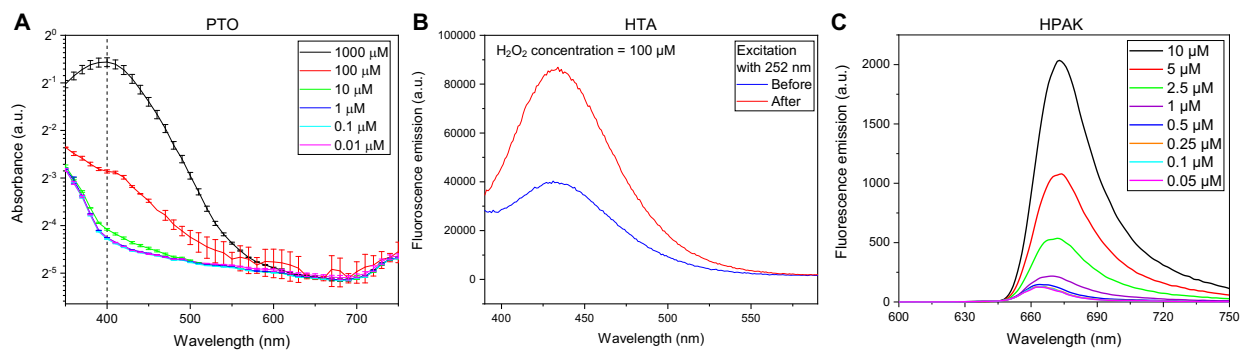

Figure S1. Comparative spectra for the following $H_2O_2$ detection methods: (A) the potassium titanium oxalate (PTO) assay, (B) the 2-hydroxyterephthalic acid (HTA) assay, and (C) the Horseradish Peroxidase assay kit (HPAK). See the Materials and Methods section for the specific methods' details.

**SI Section S1: Wetting characterization**

Contact angles of water drops on the substrates were measured using a Kruss DSA-100E system by placing 2 µl drops on the surfaces. Advancing and receding angles were recorded by adding and removing 10 µl to this drop at 0.2 µL/s rate. We used advanced software (Kruss GmbH) for image analysis by fitting tangents at the solid-liquid-vapor interface to obtain the contact angles.

Table S1. Apparent, advanced, and receding contact angles of water droplets on treated $SiO_2$/Si wafers.

| Surface treatment | Apparent contact angle, $\theta_r$ | Advancing contact angle, $\theta_A$ | Receding contact angle, $\theta_R$ |
|---|---|---|---|
| **Oxygen-plasma treated silica** | 7° | 7° | 0° |
| **FDTS-coated silica** | 105° ± 2° | 120° ± 2° | 97° ± 2° |

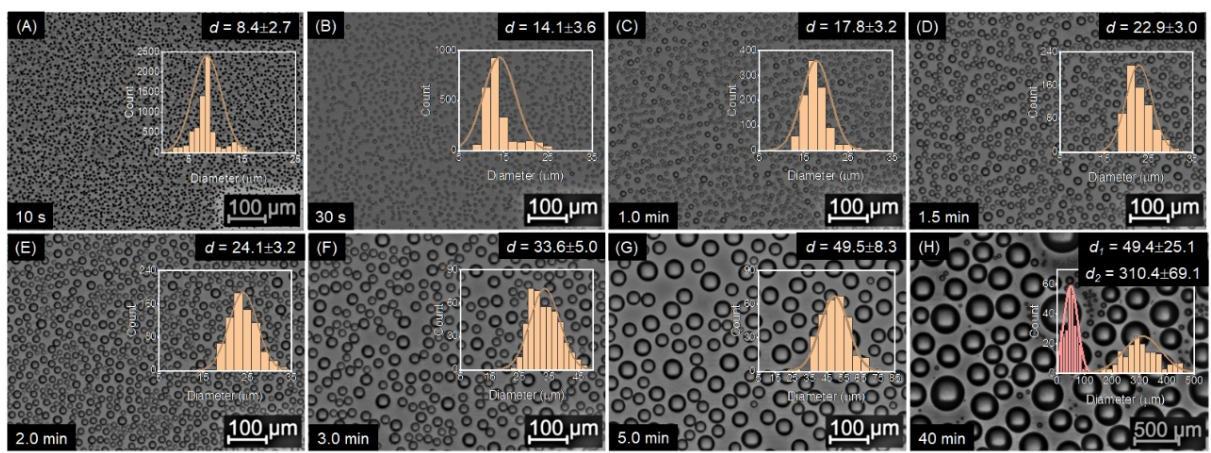

Figure S2. Representative time-dependent size distributions (mean ± standard deviation) of condensed water droplets on FDTS-coated silica surfaces maintained at 3–4 °C and 92–96% RH realized via heating water to 60 °C: (A) 10 s, (B) 30 s, (C) 1 min, (D) 1.5 min, (E) 2.0 min, (F) 3.0 min, (G) 5 min, and (F) 40 min (bimodal distribution). Note: scale bars in μm.

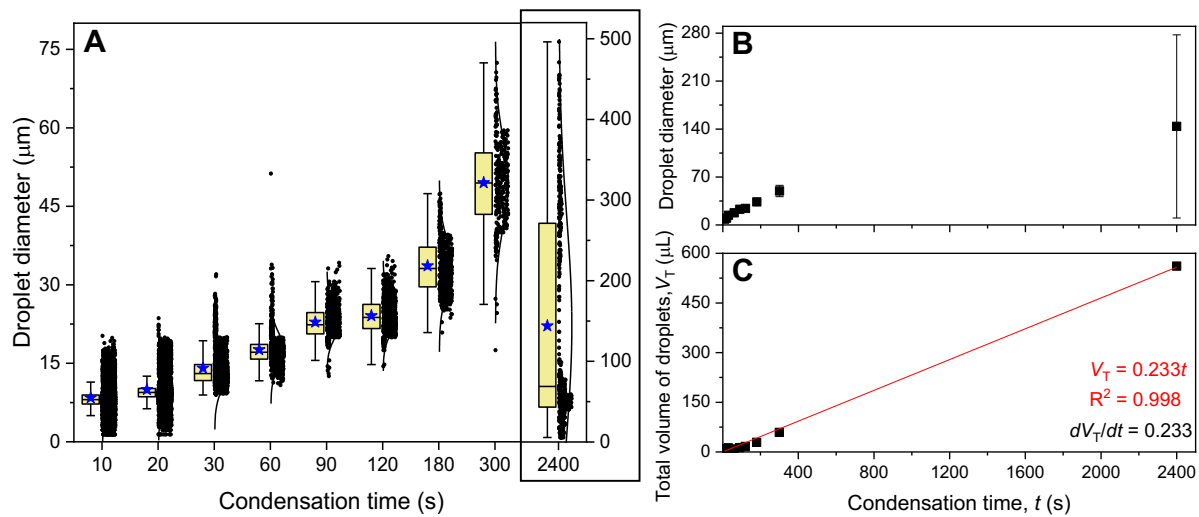

Figure S3. Representative image analysis (with ImageJ[1]) of condensed water droplets on FDTS-coated silica surfaces at 3–4 °C and 92–96% RH realized via heating water to 60 °C. (**A**) Condensed droplets' diameter distribution as a function of time (Note: blue stars correspond to the mean value; the x-axis scale is not linear and the right-hand-side y-axis is droplet diameter (µm) for $t$ = 2400 s only). (**B**) Mean droplet diameters as a function of time; error bars signify one standard deviation. (**C**) Estimated cumulative volume of the condensate over time; estimated average condensation rate: 0.23 µL/s. (Notice the bimodal droplet distribution at $t$ = 2400 s; this happens because larger droplets merge and numerous ≤ 10 µm droplets emerge in the spaces between them).

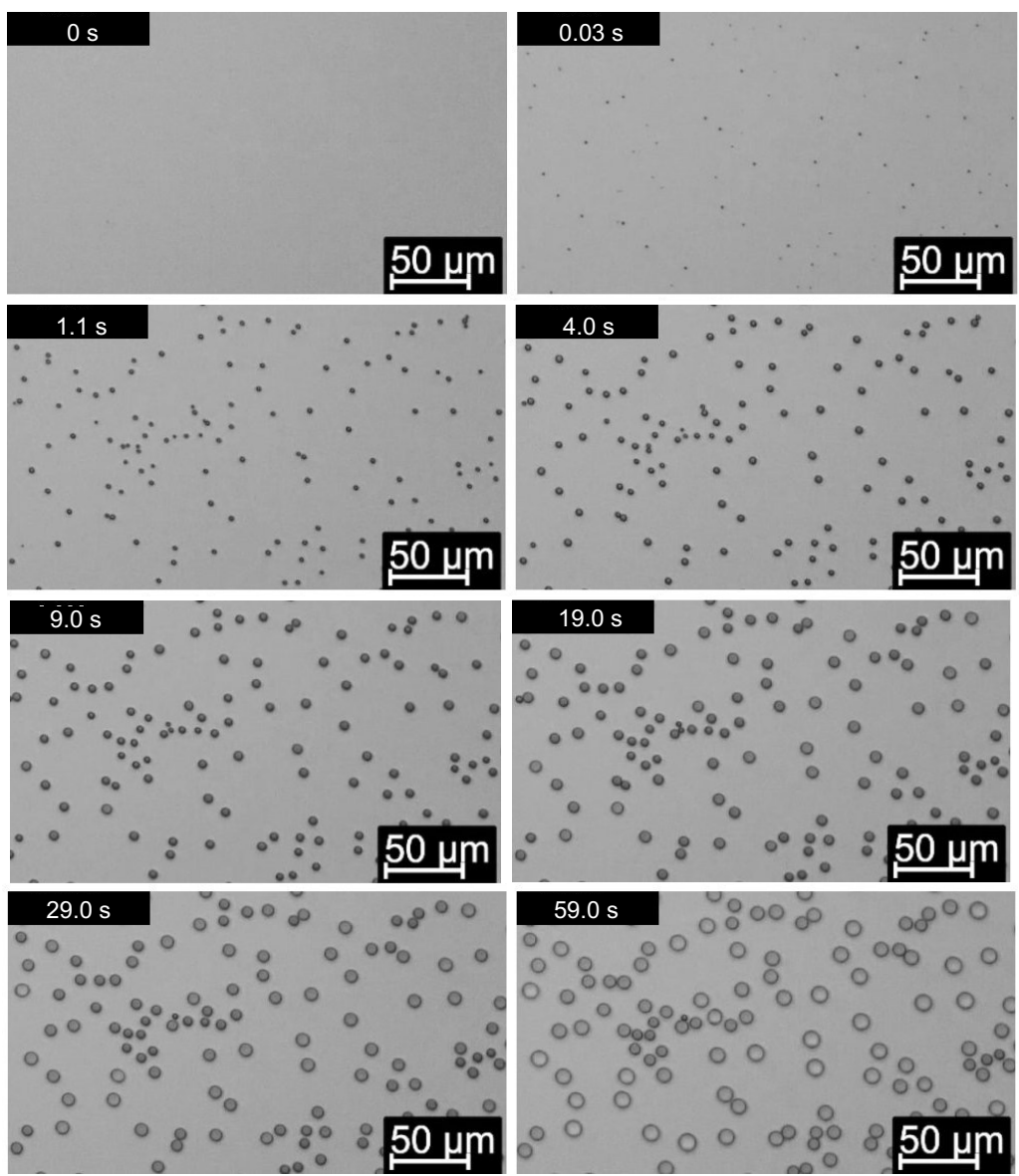

Figure S4. Representative timelapse images of water droplets condensing onto FDTS-coated silica surface maintained at 3–4 °C under 59% relative humidity and ambient air temperature of 21.4 °C.

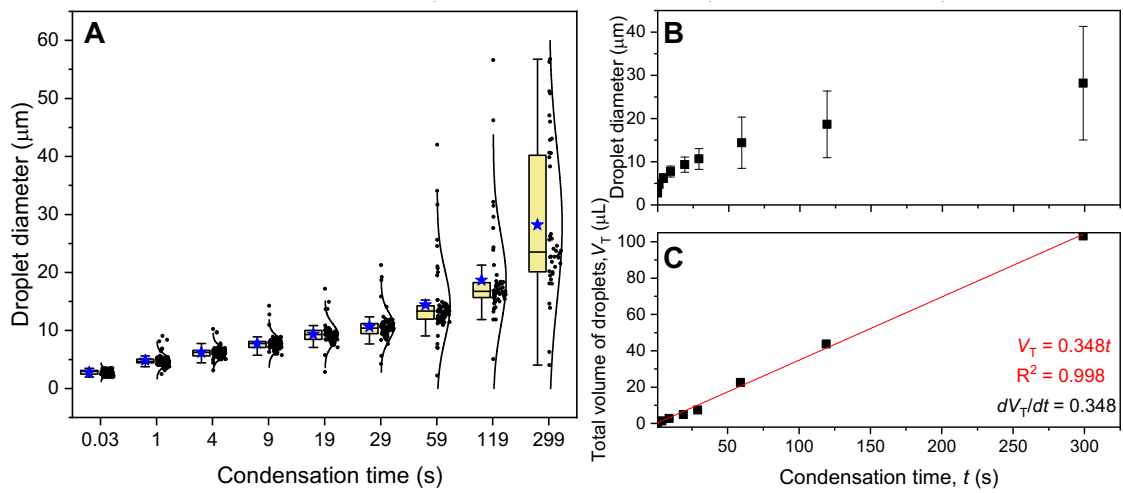

Figure S5. Representative image analysis (with ImageJ[1]) for experiment shown in Figure S4. (**A**) Condensed droplets' diameter distribution as a function of time (Note: blue stars correspond to the mean value; the x-axis scale is not linear). (**B**) Mean droplet diameters as a function of time; error bars signify one standard deviation. (**C**) Estimated cumulative volume of the condensate over time; estimated average condensation rate: 0.35 μL/s.

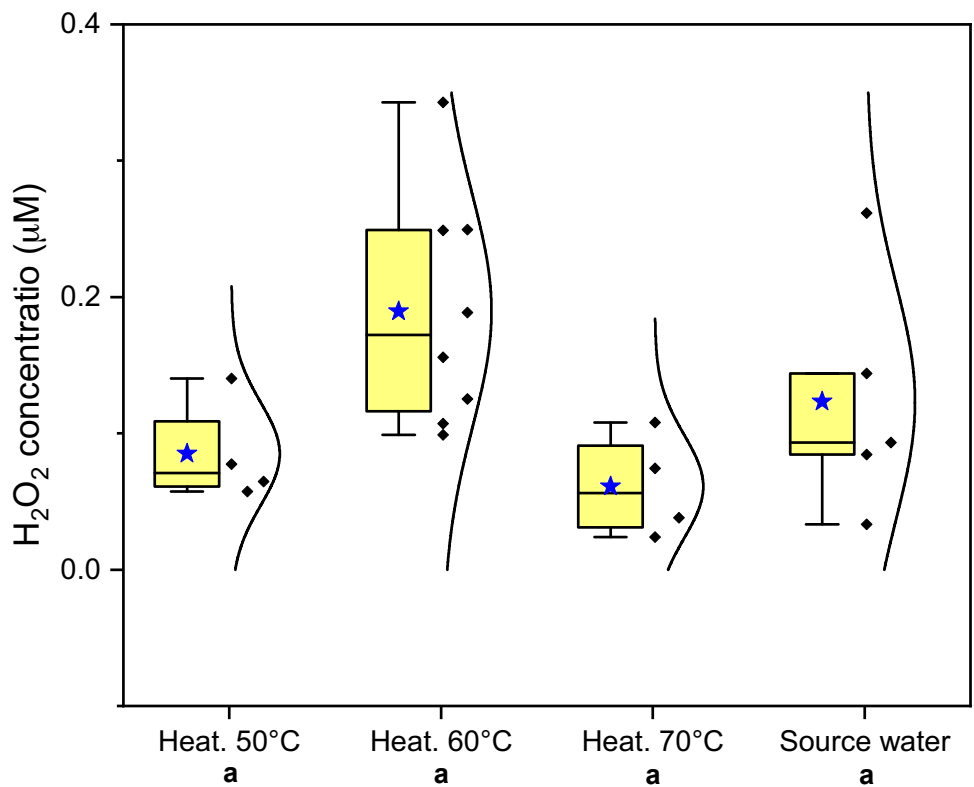

Figure S6. Comparison of $H_2O_2$ concentrations in the water droplets accumulated on hydrophobic (Phob.) and hydrophilic (Phil.) substrates after the condensation of vapor produced via (i) heating water in the temperature range of 50–70°C (Heat.). Half of the samples were condensed on FDTS-coated silica and the other half on oxygen plasma-treated silica. There was no statistically significant difference in the groups (**a**) analyzed with one way ANOVA and Tukey's test for comparison of the means ($p < 0.01$). Blue star indicates the mean value. (Note: the detection limit of the HPAK assay in our experiments was 0.25 µM).

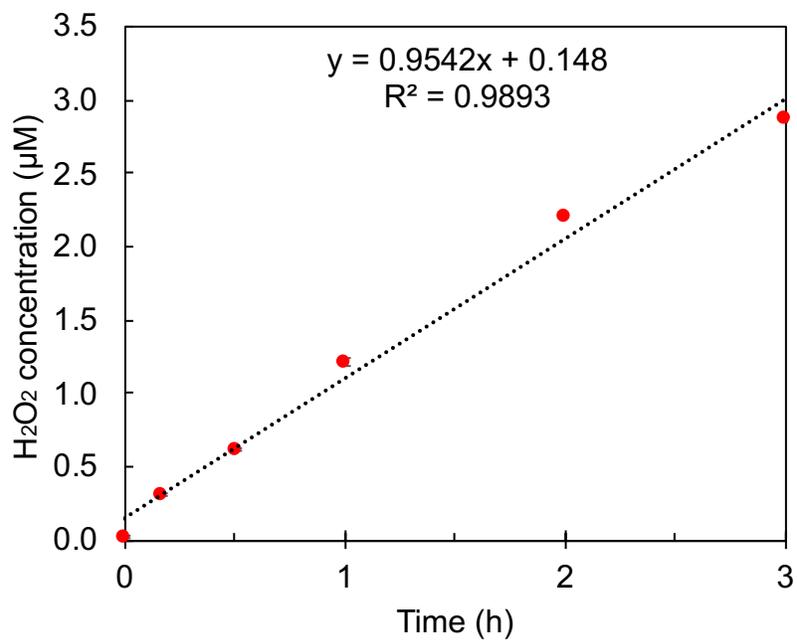

Figure S7. $H_2O_2$ production in bulk water sample using probe sonication with 30 mL of deionized water in a beaker immersed in an ice bath to reduce the evaporation.

**Section S2. HPAK can detect $H_2O_2$ despite much dilution in condensed microdroplets**

Is it possible that even HPAK with a detection limit of 0.25 µM fails to pinpoint $H_2O_2$ concentration due to dilution as the droplets' diameter exceeds 10 µm? Below, we present arguments to prove that this is not true.

First, let us assume that water microdroplets with diameter ≤ 10 µm produce ≥60 µM $H_2O_2$ in ~2 min on hydrophobic substrates, as recently demonstrated[2]. This would mandate that the water microdroplets with size distribution ≤10 µm and age 1–3 min realized via the following experiments must also contain at least 60 µM $H_2O_2$: (i) the condensation of the vapor generated from hot water (50–70 °C), (ii) the condensation of ambient water vapor (60% RH and 21 °C), and (iii) the evaporation of larger water drops on superhydrophobic substrates. However, in stark contrast to this expectation, $H_2O_2$ test strips confirmed that this is not the case. Note: as a sanity check, we tested the test strips with standard $H_2O_2$ solutions and confirmed that they reliably detect $H_2O_2$ at concentrations ≥30 µM.

Next, let us revisit the results obtained from HPAK on water microdroplets condensed onto FDTS-coated silica from the ambient air (RH 59% and 21–23 °C). Our image analysis revealed that in these experiments, microdroplets had diameters ≤10 µm up until < 20 s, after which they only got diluted. Assuming that the as-placed microdroplets exhibited the apparent advancing contact angle, $\theta_A \approx 120°$, and maintained it subsequently as it grew, we could estimate the initial total volume within a specified area to be ~5 µL (Table S1 and Figure S7C). Now, sample collection for the HPAK analysis necessitates about 400-600 µL volume, which yields a dilution factor of 80–120 (i.e., 400 µL/5 µL to 600 µL/5 µL), yielding expected $H_2O_2$ concentration to be 0.75–0.5 µM. The HPAK assay can unambiguously detect this concentration range (Figure 1C). The fact that we do not observe $H_2O_2$ concentrations in this range proves that the condensation of vapor from the ambient air or formed by heating water did not produce $H_2O_2$ even when droplet size was ≤ 10 μm (Figure 3). It should also be realized that our estimate is in fact conservative because (i) it does not take into account the <10 μm droplets that are continuously formed between the spaces left after the coalescence of the bigger droplets that putatively produce $H_2O_2$ (see the rightmost distributions in Figures 3A and S3A) and (ii) the original report actually noted up to ~115 μM $H_2O_2$ concentration, i.e., 100% higher than what we used in this argument.

# References

1. Schneider, C. A.; Rasband, W. S.; Eliceiri, K. W., NIH Image to ImageJ: 25 years of image analysis. *Nature Methods* **2012,** *9* (7), 671-675.
2. Lee, J. K.; Han, H. S.; Chaikasetsin, S.; Marron, D. P.; Waymouth, R. M.; Prinz, F. B.; Zare, R. N., Condensing water vapor to droplets generates hydrogen peroxide. *P Natl Acad Sci USA* **2020,** *117* (49), 30934-30941.